\documentclass[twocolumn,prl,superscriptaddress]{revtex4}

\usepackage{amsmath,amssymb,amsfonts}

\usepackage{graphicx}
\usepackage{lmodern}
\usepackage[T1]{fontenc}
\usepackage{color}

\newcommand{\be}{\begin{equation}}
\newcommand{\ee}{\end{equation}}

\begin{document}
\title{Instability of exotic compact objects and its implications for gravitational-wave echoes}

\author{Baoyi Chen}
\affiliation{Burke Institute of Theoretical Physics and Theoretical Astrophysics 350-17, California Institute of Technology, Pasadena, California 91125}
\author{Yanbei Chen}
\affiliation{Burke Institute of Theoretical Physics and Theoretical Astrophysics 350-17, California Institute of Technology, Pasadena, California 91125}
\author{Yiqiu Ma}
\affiliation{Burke Institute of Theoretical Physics and Theoretical Astrophysics 350-17, California Institute of Technology, Pasadena, California 91125}
\author{Ka-Lok R.\ Lo}
\affiliation{LIGO Laboratory, California Institute of Technology, Pasadena, California 91125}
\author{Ling Sun}
\affiliation{LIGO Laboratory, California Institute of Technology, Pasadena, California 91125}

\date{\today}
\begin{abstract}
Exotic compact objects (ECOs) have recently become an exciting research subject, since they are speculated to have a special response to the incident gravitational waves (GWs) that leads to GW echoes.  
We show that energy carried by GWs can easily cause the event horizon to form out of a static ECO --- leaving no echo signals towards spatial infinity.   
To show this, we use the ingoing Vaidya spacetime and take into account the back reaction due to incoming GWs. 
Demanding that an ECO does not collapse into a black hole puts an upper bound on the compactness of the ECO, at the cost of less distinct echo signals for smaller compactness.  The trade-off between echoes' detectability and distinguishability leads to a fine tuning of ECO parameters for LIGO to find distinct echoes.  
We also show that an extremely compact ECO that can survive the gravitational collapse and give rise to GW echoes might have to expand its surface in a non-causal way.
\end{abstract}

\maketitle

\noindent {\it Introduction.--} Black holes are important predictions of classical general relativity, and shown to be robust products of gravitational collapses.  The {\it event horizon}, describing the boundary within which the future null infinity cannot be reached, is the defining feature of a black hole\,\cite{Hawking1973}.  It is hoped that gravitational-wave (GW) observations can provide evidence for the event horizon.   Absence of the event horizon, as well as deviations from the Kerr geometry near the horizon, can be motivated by quantum-gravity and quantum-information considerations.  Objects whose space-time geometries mimic that of a black hole, except in the near-horizon region, have been speculated to exist, and are referred to as exotic compact objects (ECOs)~\cite{Mazur2001, Schunck2003}.  
The boundary between the Kerr and non-Kerr regions of the ECO are often placed at Planck scale above the horizon.

Pani {\it et al.} argued that GWs emitted from a gravitational collapse or a compact binary coalescence (CBC),  which results in an ECO, should be followed by {\it echoes}~\cite{Pani2009,Pani2010,Cardoso2016a, Cardoso2016b}.  GWs that propagate toward the ECO can be reflected by the ECO surface --- or penetrate through the ECO and re-merge at its surface --- and bounce back and forth between the ECO's gravitational potential barrier (at the location of the light sphere) and the ECO itself~\cite{PK2017,Nakano2017,Mark2017, WA2018,Du2018}.  GW echoes, as the smoking gun of ECOs~\cite{VP2017}, have generated enormous interest. Most notably, Abedi et al.\ claimed to have found evidence of echoes in Advanced Laser Interferometer Gravitational Wave Observatory (Advanced LIGO) data after the first few observed CBC events~\cite{Abedi2017a,Abedi2017b,Abedi2018}, and the statistical significance of these signatures was being questioned\,\cite{Ashton2016,Abedi2017b, Abedi2018,Westerweck2018,Lo2018,Nielsen:2018lkf}. Alternative techniques in searching for echoes have been proposed \cite{Westerweck2018,Tsang2018,Conklin2018,Lo2018,Nielsen:2018lkf}. In particular, many authors reported that signatures found by Abedi et al.\ were not statistically significant~\cite{Ashton2016,Westerweck2018,Lo2018,Nielsen:2018lkf}.

\begin{figure}[h] 
   \centering
   \includegraphics[width=0.6\linewidth]{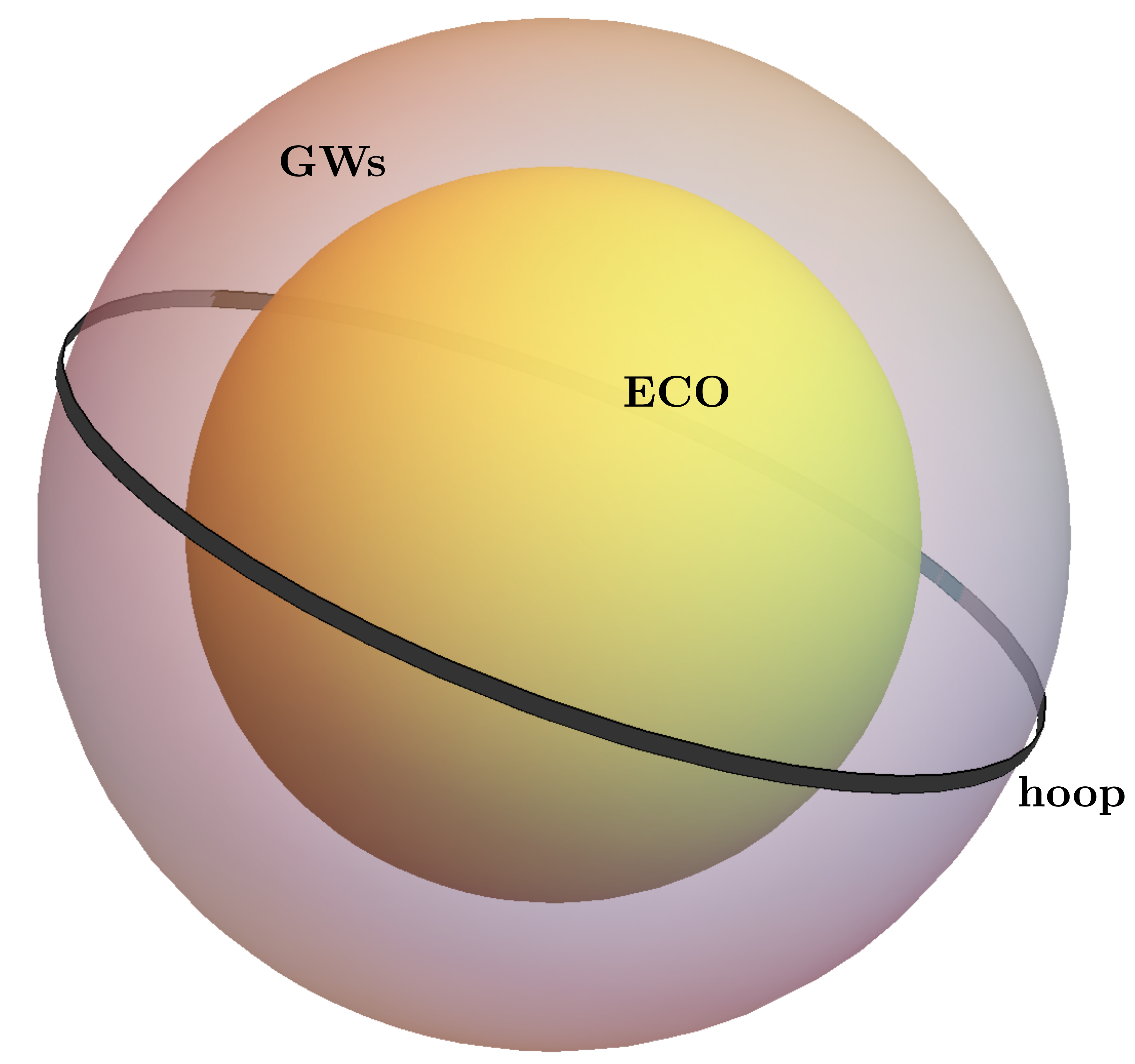} 
   \caption{A pulse of GW with energy $E$ incident on a static ECO with mass $M$.  A hoop is placed at the Schwarzschild radius $2(M+E)$.  When the spatial extent of the GWs becomes compacted within the hoop in every direction, the event horizon forms.} \label{fig:eco_static}
\end{figure}

As radiations propagate near the ECO surface, they get increasingly blue shifted (for observers with constant Schwarzschild/Boyer-Lindquist $r$, e.g., those who sit on the ECO surface); energy is also crammed into a compact region (in terms of $\Delta r)$, and we need to consider its back reaction to the space-time geometry.  As Eardley~\cite{Eardley} and Redmount~\cite{Redmount} were studying the stability of white holes and worm holes in an astrophysical environment, they concluded that the ``blue sheet'' made up by in-falling material and radiations can cause the formation of an event horizon.  Another point of view of such instability is through Thorne's hoop conjecture~~\cite{Thorne:1972ji}, as shown in Fig.~\ref{fig:eco_static}, which says a certain amount of mass/energy will collapse into a black hole when it is inside the ``hoop'' located at its own Schwarzschild radius.  
Similarly, ECOs may suffer from such instability caused by GWs and collapse into black holes. In this paper, we study the condition for ECO to remain stable and the impact of back reaction on GW echoes.

\noindent{\it Set up of the problem.--} For simplicity, we consider a spherically symmetric initial ECO with mass $M$ and areal radius $r_{\rm ECO}=2M+\epsilon$, and an incident GW packet with energy $E$. Here $\epsilon$ is a small distance that quantifies the compactness of the ECO;  $\Delta\equiv \sqrt{8 M \epsilon}$ is also used to characterize ``the spatial distance between ECO surface and the horizon''. For the ringdown phase of a CBC,  
cumulative energy emitted from the potential barrier towards the ECO, up to Schwarzschild time $t$ is given by
\begin{equation}\label{eq:qnmenergy}
E_{\rm RD}(t) \approx \alpha_{\rm H} \eta M (1-e^{-2\gamma t})\,,
\end{equation}
where $\gamma$ is the imaginary part of the quasinormal modes (QNM) frequency, $\eta=M_1M_2(M_1+M_2)^{-2}$ is the symmetric mass ratio, with $M_1$ and $M_2$ being the masses of the two objects in the binary, and the numerical factor $\alpha_\mathrm{H}$ is estimated to be around 5\%.  We define the tortoise coordinate $r_*$ from the Schwarzschild radius $r$ as $r_* = r + 2M\log\left(r/2M-1\right)$.  The {\it conventional} estimate for the time lag between the first echo and the beginning of the ringdown signal is given by
\begin{equation}
\label{dtconv}
    \Delta t^{\rm conv}_{\rm echo} = 2|r_*^{\rm LR} -r_*^{\rm ECO}| \approx 2M+ 4M \log (M/\epsilon).
    \end{equation}
where LR stands for the light ring at $r^{\rm LR}=3M$, and $r^{\rm ECO}_*$ is the tortoise coordinate for $r_{\rm ECO}$ \footnote{In some existing literature, the time lag between the first echo and the beginning of the ringdown signal is denoted by $t_{\rm echo}$, and notation $\Delta t_{\rm echo}$ is used to denote the time lag between two echoes.}.

\noindent {\it Estimates based on the hoop conjecture.--} 
According to the \textit{hoop conjecture}~\cite{Thorne:1972ji}, the event horizon forms when a certain amount of mass gets compacted within its own Schwarzschild radius.  This corresponds to a zeroth-order estimate on the effect of the incoming energy towards an ECO, in the sense that we neglect the back reactions of the GWs to the ECO Schwarzchild spacetime.  
As shown in Fig.~\ref{fig:eco_static}, at the instant when all incoming energy gets compacted within the ``hoop'', the event horizon forms. 
If we consider the null packet carrying the ringdown energy~\eqref{eq:qnmenergy}, to prevent horizon formation, the location of the ECO surface must satisfy
\be
r_{\rm ECO} - 2M  > 0.04\eta M ({M\gamma}/{0.1})({\alpha_{\rm H}}/{0.05})\,.
\ee
This means, stable, static ECOs cannot be very compact ---$\epsilon$ or $\Delta$ are far from Planck scale.

\begin{figure}[t] 
   \centering
   \includegraphics[width=0.85\linewidth]{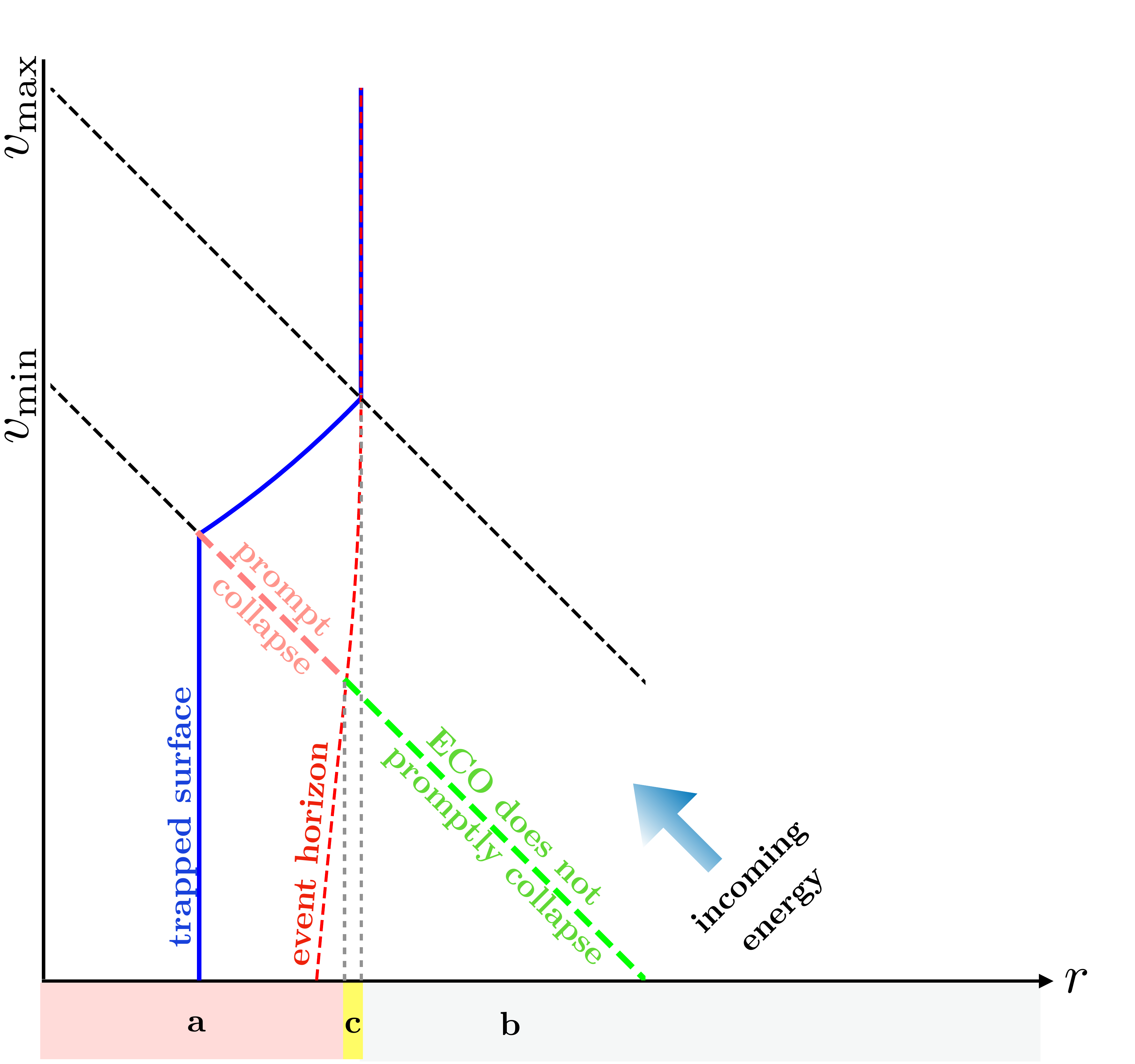} 
   \caption{A Vaidya spacetime with an incoming null packet spatially bounded by the black dashed line.  The trapped surface evolves along the blue solid line. The event horizon evolves along the red dashed line, and coincides with the trapped surface after all energy goes into the horizon. Any ECO with its surface crosses the pink line will promptly collapse, while those crosses the green line does not.
   Static ECOs can then be divided into three different types (a), (b) and (c) separated by the gray dashed lines (not world lines), and are discussed in detail in the main text.} \label{fig:horizon}
\end{figure}

\noindent{\it In-going Vaidya Spacetime.--}  To approach a more accurate study of the back reaction of the incoming GWs, we consider a Vaidya geometry, 
\begin{equation}
ds^2 =-\left[1-2M(v)/r\right] dv^2  +2 dr dv + r^2d\Omega^2\,,
\end{equation}
where $v$ is the advanced time, and $M(v)$ is the total gravitational mass that has entered the space-time up to $v$. The Vaidya geometry, as an approximation, describes a spherically symmetric space-time with (gravitating) in-falling dust along radial null rays, but does not capture the fact that GW energy is not spherically symmetrically distributed --- nor does it capture GW oscillations.  For incident GWs during $v_{\rm min} < v < v_{\rm max}$, we have 
\begin{equation}
    M(v)=M_{\rm min}+ E(v)\,,
\end{equation}
where $E(v)$ is total GW energy that has entered since $v_{\rm min}$.  $M$ grows from $M_{\rm min}$ to $M_{\rm max} \equiv M_{\rm min}+E_{\rm tot}$. During the process of incoming GWs, as shown in Fig.~\ref{fig:horizon}, the apparent horizon (AH) traces the total energy content, and is located at $r_{\rm AH}=2M(v)$. 
The event horizon (EH), on the other hand, follows out-going radial null geodesics, parameterized by $r(v)$, which satisfies
\begin{equation}
2{dr}/{dv}=1-{2M(v)}/{r(v)}\,.
\end{equation}
We also need to impose a {\it final condition} of $r_{\rm EH}(v_{\rm max}) = 2M_{\rm max}$. Assuming $\dot M \ll 1$, writing $r_{\rm EH}(v)=2M(v)+\delta(v)$ with $\delta \ll M$, we have
$
\dot\delta-\delta/(4 M)=-2\dot{M}
$, 
and the solution is given by
\begin{equation}
\delta (v)=2\int^{v_{\rm max}}_vdv' \dot{M}(v')\mathrm{exp}\left[-\int^{v'}_{v}dv''\frac{1}{4M(v'')}\right]\,.
\end{equation}
The required final condition for $\delta$, as well as the dependency of $\delta(v)$ on $\dot M(v')$ at $v'>v$, embody the {\it teleological nature} of the EH: the location of the EH right now is determined by what shall happen in the future.

For the ringdown of a CBC, we substitute $E(v)=E_{\rm RD}(v-v_{\rm min})$ into the solution and obtain 
\begin{equation}
    \delta (v) = \frac{16\eta\alpha_{\rm H} \eta M (M\gamma)}{1+8M\gamma} e^{-2(v-v_{\rm min})} \equiv \epsilon_{\rm th}  e^{-2(v-v_{\rm min})}\,. 
\end{equation}
Next, we use Vaidya spacetime only as the {\it exterior} of the ECO, and 
consider two scenarios: (i) GW-induced collapse of a static ECO and (ii) an ECO with expanding surface. 

\begin{figure*}[t] 
   \centering
   \includegraphics[width=\textwidth]{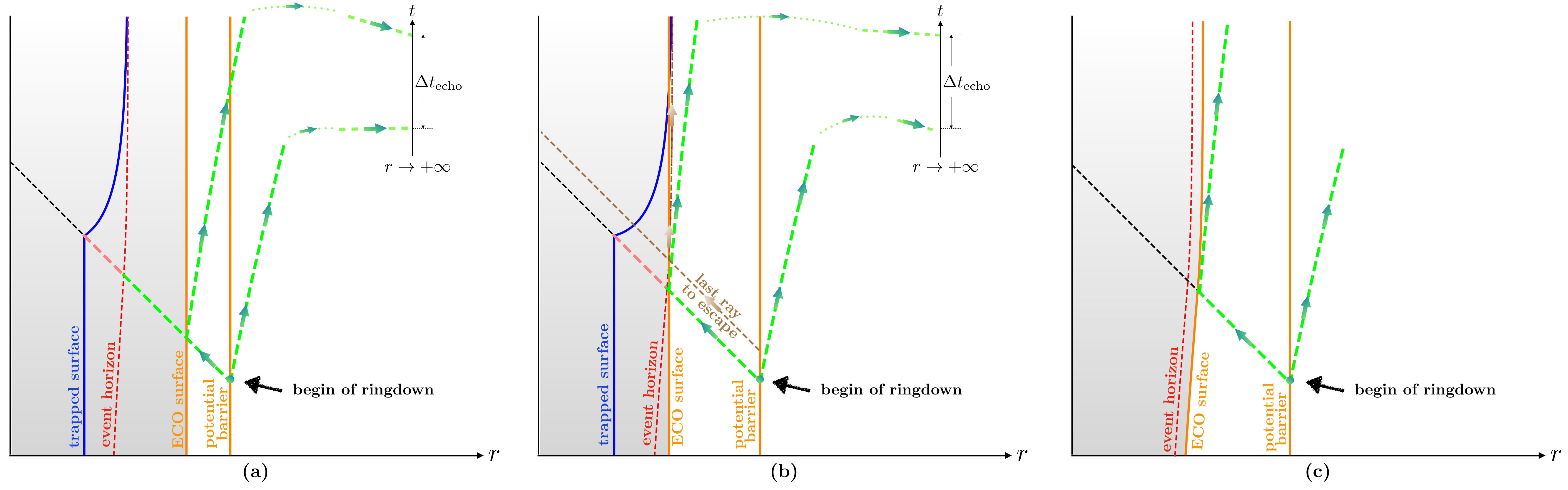} 
   \caption{(a) A static ECO scenario where the ECO surface is outside the event horizon and, which can give rise to GW echoes.  The spatially nearest incoming ray, denoted by the green dashed line, reflects at the potential barrier and the ECO surface, leading to a time delay $\Delta t_{\mathrm{echo}}$ between the main wave and the echo as observed at spatial infinity. (b) A static ECO scenario where part of the incoming GW energy gets reflected at the ECO surface and gives rise to echoes until ``the last ray to escape''. The subsequent echoes are highly redshifted due to the formation of event horizon, leading to a weak echo signal. (c) Spherically symmetric ECO absorbing GWs and expanding in radius. The event horizon grows along the red dashed line. If the ECO surface always remains outside the event horizon, incoming rays can lead to GW echoes, as shown by the green dashed lines. }\label{fig:eco_three_type}
\end{figure*}

\noindent {\it Back reaction: static ECO with future incoming pulse.--} For static ECOs with radius $r=r_{\rm ECO}$ and initial mass $M_{\rm min}$, we can divide them into three types as in Fig.~\ref{fig:horizon}: 

(a) For $r_{\rm ECO}<2M_{\rm min}+\epsilon_{\rm th}$, the ECO will promptly collapse and there will be no GW echoes, since the first incoming ray reaches the ECO inside the EH.  

(b) For $r > 2 M_{\rm max}$, the ECO does not collapse, and conventional echoes (generally more than one) will form, as individually shown in Fig.\,\ref{fig:eco_three_type}a.

(c) For $2M_{\rm min}+\epsilon_{\rm th}<r_{\rm ECO}<2M_{\rm max}$, the ECO enters the EH (hence collapses) during the incoming GW pulse.  Only one echo, with reduced magnitude, could form, as individually shown in Fig.~\ref{fig:eco_three_type}b.

The magnitude of $\epsilon_{\rm th}$ indicates that static ECOs that can produce echoes will deviate from a black hole at a distance far from Planck scale above the horizon. In terms of compactness, for $M\gamma\approx 0.1$, we have
\begin{equation}
    \epsilon_{\rm th}/(2M) =5.6\times10^{-3} (\alpha_{\rm H}/0.05) (\eta/0.25).
\end{equation}
For a comparable-mass binary (e.g., $\eta = 0.25$), this corresponds to a moderate bound on the compactness; for extreme mass-ratio inspirals (EMRIs) with $\eta=10^{-7}$, compactness $\epsilon_{\rm th}/(2M)$ reaches $4\times 10^{-9}$ for typical values of $\alpha$ and $M\gamma$. In terms of proper distance above the horizon, we have, $M\gamma\approx 0.1$,
\begin{equation}
    \Delta_{\rm th}= 
    8\sqrt{\frac{2\alpha_{\rm H} \eta  M^3\gamma}{1+8M\eta}}\approx 0.6 \sqrt{M_1 M_2} \sqrt{\alpha_{\rm H}/0.05}\,,
\end{equation}
For stellar mass CBC, $\Delta_{\rm th}$ is at least kilometer-scale, far from Planck scale.

\noindent{{\it Back reaction: expanding ECO.--}} If the ECO has an expanding surface, as shown in Fig.~\ref{fig:eco_three_type}c, its increasing radius could {\it in principle} keep up with the influx of GW energy, so that the horizon will not form. 
For example, {\it if} 
\be
\label{expandingeco}
r_{\rm ECO}(v) =r_{\rm EH}(v)+\epsilon =    2M(v)+\delta(v)+\epsilon\,,
\ee 
with a constant positive $\epsilon$ which can be aribitrarily small, ECO surface will be time-like. We need to emphasize that the expansion trajectory~\eqref{expandingeco}, although being time-like, is {\it teleological} in nature, since its rate of expansion must be determined by the {\it future} in-going GW flux.  In other words, internal physics of the ECO must know how much energy is going to come in the future, and adjust the ECO radius accordingly, before the waves arrive.

\noindent{\it  Implications for GW-echo phenomenology.--} 
The back reaction of incoming GWs substantially affects the phenomenology of GW echoes by imposing constraints on $\Delta t_{\rm echo}$, which is the first echo's time lag behind the main wave.  When $\Delta t_{\rm echo}$ is comparable to the ringdown time scale $1/\gamma$, the echoes will interfere with the main wave~\cite{Mark2017}, giving rise to less distinct echo signals.  To better illustrate this, we define a ratio $R$ between these two time scales via $R \equiv  \gamma\Delta t_{\rm echo}$. Then the echo is separated form the main wave when $R \ge 1$.

\begin{figure}[h]
    \centering
    \includegraphics[width=\linewidth]{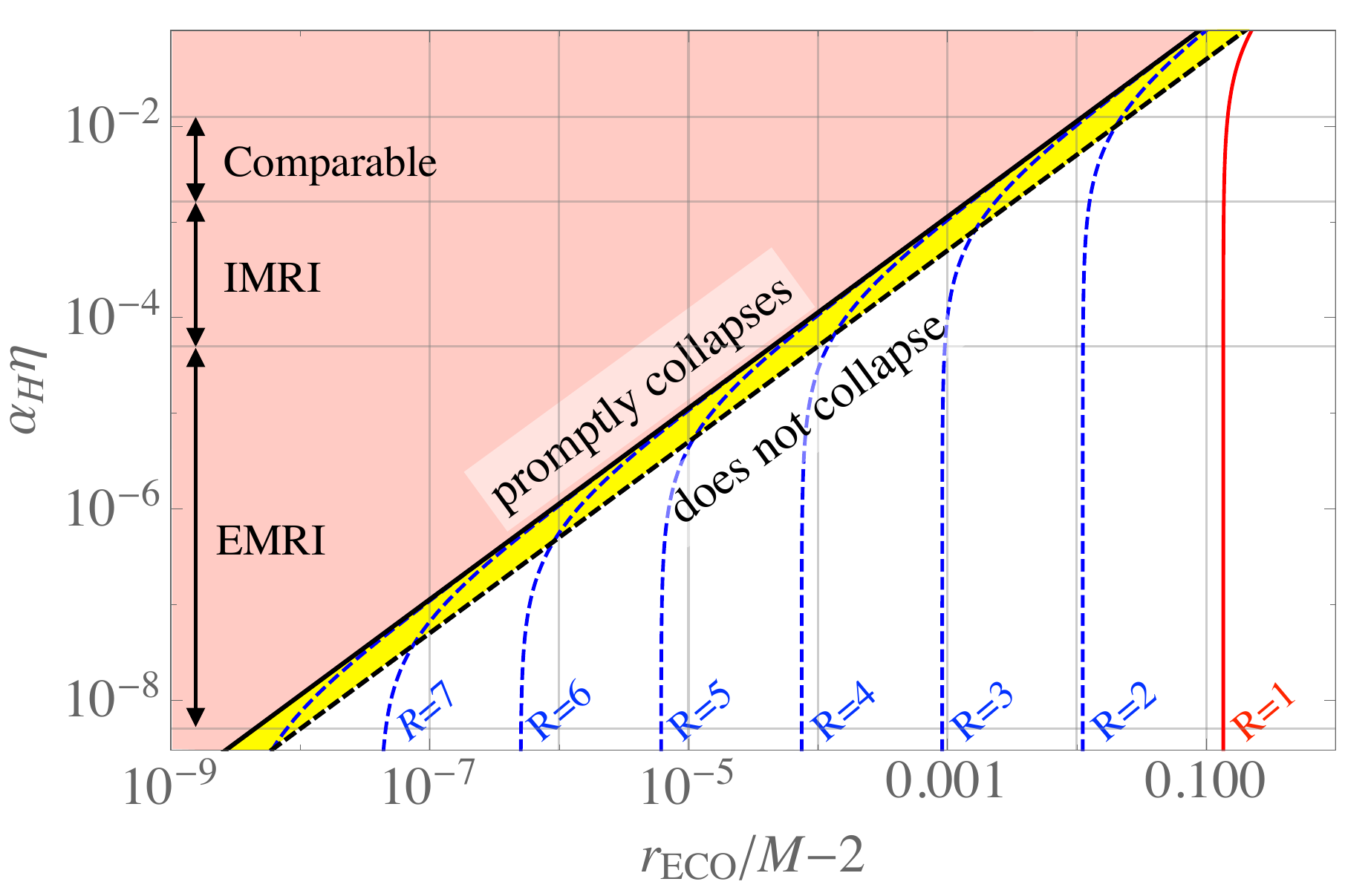}
    \caption{Contour plot for $R$ as function of $r_{\rm ECO}/M-2$ and $\alpha_{\rm H}\eta$. Regions in red, white, and yellow indicate types (a), (b), and (c), respectively. Along the vertical axis, the ranges indicate comparable, intermediate, and extreme mass ratio inspirals.}
    \label{fig:contour}
\end{figure}

For static ECOs of type (a), which promptly collapse, there are no echoes. For types (b) [Fig.~\ref{fig:eco_three_type}a] and (c) [Fig.~\ref{fig:eco_three_type}b], the ratio $R$ can be obtained from 
\begin{equation}
\label{eqtune}
    r_{\rm ECO}/M -2 =\epsilon_{\rm th}/M +\exp[-R/(4 M\gamma)+1/2]\,.
\end{equation}
In Fig.~\ref{fig:contour}, we plot the contour of $R$ as function of $r_{\rm ECO}/M-2$ and $\alpha_{\rm H}\eta$.  
For type (b), echo arises from the entire duration of incoming GW, there will be {\it subsequent echoes}, when GW reflected from the ECO further travel back and forth between the potential barrier and the ECO.  For type (c), echo arises only from the first part of the incoming GW, up till the ``last ray to escape'' shown in Fig.~\ref{fig:eco_three_type}b, and {\it there can only be one echo}. The reflected GW will first oscillate and then be ``frozen" due to ECO gravitational collapse. Since the low-frequency component of the reflected GW can not propagate to infinity due to the filtering of the frequency-dependent potential barrier, the distant observer will just see a weakened QNM waveform.

For CBCs observable by LIGO, we choose, for instance, $\eta=0.25$, $\alpha_\mathrm{H}=0.05$ and $M \gamma=0.1$, which are consistent with GW150914 \cite{GW150914TGR}. We then have $\epsilon_{\rm th}=0.011M$. Type (b) ECOs  should have $r_{\rm ECO} > 2.025M$. In particular, for $2.025M<r_{\rm ECO}<2.15M$, we have $1<  R <1.9$.  For type (c) ECOs, we could have much larger $R$, but that would correspond to a small region in parameter space. 
Previously, one could have argued that exotic physics could create static ECOs that have $r_{\rm ECO}/M$ very close to 2.  However, to have a moderately large $R$, Eq.~\eqref{eqtune} requires that $r_{\rm ECO}/M$ be exponentially close to $2+\epsilon_{\rm th}$, with $\epsilon_{\rm th}$ depending on the incoming GW.  This seems a fine tuning for static ECOs which is very unlikely to happen.

For EMRIs targeted by Laser Interferometer Space Antenna (LISA), less incoming GWs allows more compact ECOs to be probed. Let us consider $\alpha_{\mathrm{H}}=0.05$, $\eta=10^{-7}$ and $M\gamma=0.1$, 
and use the boundary between type (b) and type (c) as the typical ECO compactness. In this case, we will have \mbox{$R^{\rm typical} \approx 7.8$}, which corresponds to a distinct echo signal.

In the case of expanding ECOs with $r_{\rm ECO}=2M(v)+\epsilon$, under the approximation that $\dot{M}(v)\ll 1$, $\Delta t_{\rm echo}$ is given by 
$\Delta t^{\rm expand}_{\rm echo} = 2M+4M\log[{M}/{\epsilon}]$, which is equal to $\Delta t^{\rm conv}_{\rm echo}$.
Here we see that a teleologically expanding ECO has the same phenomenology proposed in existing literature.

\noindent {\it Discussions.--} 
As an attempt to use simplified analytical solutions to capture features of a highly complex space-time geometry, governed by yet unknown physics, our work has several limitations. 
(i) We have focused on the echo of {\it reflective type}, i.e, the echo generated by the reflection of the main wave on the ECO surface. For those echoes of {\it transmissive type}, i.e, the echo generated by the GW penetration into and re-emerge out of the ECO surface, the delay of echo signal depends on the specific ECO model.
(ii) Our Vaidya spacetime model only captures the back reaction of the ingoing GW flux, while in reality the reflected outgoing GWs also gravitate. The back reaction of the reflected waves may have qualitatively significant effects on the echoes when the surface reflectivity is large. 
(iii) We have not attempted to describe what happens as the other object impacts the ECO, which takes place roughly at the same time as the ringdown wave starts to impinge on the final compact object.  Finally, while the teleological response necessary for the expanding ECO sounds unnatural, it might arise due to  {\it non-local} interactions that were speculated to exist near the event horizon~\cite{Giddings}; we also note that the final-state projection model~\cite{HM2004,LP2014} may also be regarded as  teleological in nature.

\noindent {\it Acknowledgements.--} We thank Davide Gerosa, Sizheng Ma, and Kip Thorne for discussions.  Research of BC, YC and YM are supported by the National Science Foundation through Grants PHY-1708212 and PHY-1708212, the Brinson Foundation, and the Simons Foundation (Award Number 568762).  Research of LS and KLRL are supported by the LIGO Laboratory. LIGO was constructed by the California Institute of Technology and Massachusetts Institute of Technology with funding from the National Science Foundation, and operates under cooperative agreement PHY--0757058. Advanced LIGO was built under award PHY--0823459. KLRL would also like to gratefully acknowledge the support from the Croucher Foundation in Hong Kong. This paper carries LIGO Document Number LIGO-P1900056.

\end{document}